\documentstyle[11pt,amsfonts]{article}
\renewcommand{\theequation}{\arabic{equation}}
\newcommand{\be}{\begin{equation}}
\newcommand{\ee}{\end{equation}}
\newcommand{\bea}{\begin{array}}
\newcommand{\ea}{\end{array}}
\newcommand{\beqa}{\begin{eqnarray}}
\newcommand{\eeqa}{\end{eqnarray}}
\newcommand{\bean}{\begin{eqnarray*}}
\newcommand{\eean}{\end{eqnarray*}}

\def\up#1{\leavevmode \raise.16ex\hbox{#1}}

\newcommand{\gapproxeq}{\lower
 .7ex\hbox{$\;\stackrel{\textstyle >}{\sim}\;$}}
\newcommand{\lapproxeq}{\lower .7ex\hbox{$\;\stackrel
{\textstyle <}{\sim}\;$}}

\renewcommand{\theequation}{\thesection.\arabic{equation}}

\newcounter{appendice}
\newcommand{\appendice}
{
\setcounter{equation}{0}
\renewcommand{\theequation}{\Alph{appendice}.\arabic{equation}}
\addtocounter{appendice}{1}

{\Large{\bf  Appendix \Alph{appendice}}}
}

\def\thebibliography#1{{\bf REFERENCES\markboth
 {REFERENCES}{REFERENCES}}\list
 {[\arabic{enumi}]}{\settowidth\labelwidth{[#1]}\leftmargin\labelwidth
 \advance\leftmargin\labelsep
 \usecounter{enumi}}
 \def\newblock{\hskip .11em plus .33em minus -.07em}
 \sloppy
 \sfcode`\.=1000\relax}

\def\BI{{\rm 1\!l}}

\begin{document}
\centerline{ \LARGE Noncommutative Static Strings from  Matrix Models}

\vskip 2cm

\centerline{ A. Stern\footnote{astern@ua.edu}   }

\vskip 1cm
\begin{center}
  { Department of Physics, University of Alabama,\\ Tuscaloosa,
Alabama 35487, USA\\}

\end{center}
\vskip 2cm

\vspace*{5mm} 

\normalsize
\centerline{\bf ABSTRACT}

We examine the  noncommutative cylinder  solution to a matrix model with a Minkowski background metric.   It can be regarded as the noncommutative analogue of a static circular string.  Perturbations about the solution yield   a tachyonic scalar field (and an additional tachyonic fermion in the full supersymmetric version of the  model) in the commutative limit.  The  tachyonic mode is attributed to the fact that the circular  string is unstable under uniform adiabatic deformations. We obtain a stabilizing term which when added to the matrix model  removes the tachyonic mass.

\bigskip
\bigskip

\newpage

\section{Introduction}
 
Matrix models were introduced  to capture nonperturbative aspects of string theory.\cite{Ishibashi:1996xs},\cite{Banks:1996vh}                                                     Space-time geometry,   field theory and even gravity can dynamically emerge from matrix models.\cite{Aoki:1998vn},\cite{Steinacker:2010rh}  The resulting space-time geometry is a feature of the solutions to  matrix models.  In the generic case, it is  noncommutative, with a straightforward commutative limit.  A well studied example  is the fuzzy sphere, which has finite dimensional representations and a commutative limit corresponding to the sphere.\cite{Madore:1991bw},\cite{Grosse:1994ed},\cite{CarowWatamura:1998jn},\cite{Alexanian:2000uz},\cite{Dolan:2001gn},\cite{Balachandran:2005ew}   Perturbations about this solution yields a gauge theory on the fuzzy sphere.\cite{Iso:2001mg} The background metric in this case is Euclidean. Counterparts in Minkowski space-time are the fuzzy de Sitter and anti-de Sitter solutions, which  have  infinite dimensional representations and have been studied in \cite{Ho:2000fy},\cite{Jurman:2013ota}.  Here we exam  the   noncommutative cylinder  solution, which also has infinite dimensional representations.  This solution, whose commutative limit is the cylinder (or  static circular string), holds in either Euclidean or Minkowski space, although we specialize to the latter where the noncompact direction corresponds to the time. The noncommutative cylinder has novel features, such as a discrete spectrum for the time operator,\cite{Chaichian:2000ia},\cite{Balachandran:2004yh},\cite{Steinacker}  and it  appears in a noncommutative version of Ba\~nados, Teitelboim, Zanelli (BTZ) black hole geometry.\cite{Dolan:2006hv} Here we shall show that the field theory resulting from perturbations about the noncommutative  cylinder solution   describes a tachyonic scalar, and this is due to the fact that the static circular string is unstable under uniform adiabatic deformations. We  propose the addition of a term to the matrix model which removes this instability.

 The matrix model setting for this article is a three-dimensional version of the Ishibashi, Kawai, Kitazawa and Tsuchiya (IKKT)  model, with  a cubic  term  included in the bosonic action. It is identical to the matrix model used  in  \cite{Jurman:2013ota}.  The target space for the  bosonic sector is thus  spanned by three infinite dimensional Hermitean matrices, which transform covariantly under  the action of the $2+1$ Lorentz group. The fuzzy de Sitter and anti-de Sitter solutions of \cite{Jurman:2013ota} are preserved under the homogeneous  symmetry transformations of the target space.  On the other hand, the noncommutative  cylinder is an example of a symmetry breaking solution, where $2+1 $ Lorentz symmetry of the target space is broken to  $SO(2)\times \tilde T$. $SO(2)$ correspond to rotations in the plane, while $\tilde T$ refers to discrete time translations.  Perturbations about the solution lead to a nontrival noncommutative field theory, which in the commutative limit describes a scalar field coupled to a $U(1)$ gauge field on the cylinder.  The gauge field has no dynamics, since  it lives on a two dimensional space-time, and instead induces a tachyonic mass to the scalar field.  As stated above, this can be traced back  to the instability of the classical string solution.  We show that the instability is cured with the addition of explicit symmetry breaking terms to the matrix model action.  The full supersymmetric theory also yields a tachyonic fermionic field in the commutative limit. This tachyonc instability is also easily eliminated with a supersymmetry breaking contribution to the action.

The outline of this article is the following:  We begin with  closed classical string solutions in section 2. The matrix model analogues are discussed in section 3, while perturbations about the noncommutative cylinder solution are given in section 4.  We consider the generalization to $U(N)$ gauge theory and the fermionic sector of the ${\cal N}=1$ supersymmetric extension  in section 5.  The stabilizing term is discussed in section 6, and a topological model is considered in section 7.  Some concluding remarks are made in section 8.  We review the Moyal-Weyl star product on a noncommutative cylinder in appendix A, while our spinor conventions are given in  appendix B.

\section{Static string solution}
\setcounter{equation}{0}

We start with a closed  string in $2+1$ Minkowski space-time, and denote the embedding coordinates by $x^\mu,\;\mu=0,1,2$.    $\xi^a,\; a=0,1$, will parametrize the string world sheet $\Sigma=R^1\times S^1$, where $\xi^0$ ($-\infty<\xi^0<\infty$) is the time coordinate and $\xi^1$  ($-\pi\le \xi^1<\pi$) the space coordinate. For the  action we take  the standard  Nambu-Goto form plus an additional interaction term which we denote by $S_{NS}$
\be S_{string}=-{\cal T}\int d^2\xi \sqrt{-{\tt g}} +S_{NS}\;,\label{clstactn}\ee where ${\sqrt{\tt g}}$ is the determinant of the induced metric ${\tt g}_{ab}=\partial_ax^\mu\partial_b x_\mu$ on $\Sigma$,  $\partial_a=\frac \partial{\partial \xi^a}$ and the constant ${\cal T}$ denotes the string tension. Our convention for the background space-time metric is $\eta=$diag$(-1,1,1)$.  The  second term  in (\ref{clstactn}) is
\be  S_{NS}=-\frac{\cal  T}{6\rho}\int  \epsilon_{\mu\nu\rho}x^\mu dx^\nu\wedge dx^\rho\;, \label{NScplng}\ee
where the constant $\rho$  has units of  length. It can be regarded as a coupling to a Neveu-Schwarz field of the from $B_{\mu\nu}\propto \epsilon_{\mu\nu\lambda}x^\lambda$. Both terms in the action (\ref{clstactn}) are reparametrization invariant, and respect the Poincar\'e symmetry of $2+1$ Minkowski space-time.

Upon extremizing the total action with respect to variations in coordinates $x^\mu$ one gets
\be  \Delta x_\mu +\frac 1{\rho}n_\mu=0\;,\label{cleom}\ee where $\Delta =-\frac 1{\sqrt{-{\tt g}}}\partial_a\sqrt{-{\tt g}} {\tt g}^{ab}\partial_b$ is the Laplace-Beltrami operator on  $\Sigma$, ${\tt g}^{ab}$ denotes the components of the inverse induced metric, ${\tt g}^{ab}{\tt  g}_{bc}=\delta ^a_c$, and $n_\mu=\frac 1{2{\sqrt{-{\tt g}}}}\epsilon^{ab}\epsilon_{\mu\nu\rho}\partial _a x^\nu\partial_b x^\rho$ is a space-like unit vector normal to $\Sigma$.  [Our conventions are $\epsilon^{01}=\epsilon_{012}=1$.]  The equations of motion imply the existence of a conserved current $p^{a}_\mu$ on the world sheet
\be \partial_ap^{a}_\mu=0 \;,\qquad\quad p^{a}_\mu=-{\cal T}\sqrt{-{\tt g}} {\tt g}^{ab}\partial_b x_\mu+\frac {\cal T}{2\rho}\epsilon^{ab}\epsilon_{\mu\nu\rho}\, x^\nu\partial_b x^\rho\label{panu}
\ee  From $p^{a}_\mu$ one can then construct the  stress-energy tensor in the three-dimensional  embedding space, \be T^{\mu\nu}(y)=\int d^2\xi\, p^{ a\nu}\partial_ax^\mu \,\delta^3(y-x(\xi))\;,\label{strsnrg}\ee satisfying $\frac\partial{\partial y^\mu} T^{\mu\nu}(y)=0$.

 The equations of motion (\ref{cleom}) are satisfied by\cite{Jurman:2013ota}  
\be \pmatrix{x^0\cr x^1\cr x^2}=2\rho \pmatrix{\tan \xi^0\cr  \sec\xi^0\cos \xi^1\cr \sec\xi^0 \sin\xi^1}\;,\label{adstwo}\ee leading to 
\be
-(x^0)^2+(x^1)^2+(x^2)^2=4\rho^2\ee  It corresponds to two-dimensional de Sitter space $dS^2$.\footnote{$AdS^2$ solutions result if the background space-time metric is  changed to $\eta=$diag$(-1,-1,1)$, but we will not consider that possibility. Our focus instead will be on the static solution (\ref{sss}).} 
 Here we must restrict the range of $\xi^0$ to $-\frac\pi 2\le \xi^0\le\frac\pi 2$. This is a zero energy  configuration.   For this we define the energy of the string at a given time $y^0$ as  $\int dy^1dy^2\, T^{00}(y) $.  Moreover, all components of  the stress-energy tensor vanish in this case.  (\ref{adstwo})  is thus degenerate with the vacuum solution.

Another solution to the system, which is of interest in this article, is the  closed static string 
\be \pmatrix{x^0\cr x^1\cr x^2}=\pmatrix{\xi^0\cr \rho \cos \xi^1\cr \rho \sin\xi^1}\;,\label{sss}\ee where the world sheet is a cylinder of   radius  $\rho$, 
\be (x^1)^2+(x^2)^2=\rho^2\ee
The energy of this  string configuration is nonzero; More specifically, the energy equals the string tension times one-half the circumference, $\int dy^1dy^2\, T^{00}(y)=\pi \rho {\cal T}$.

Next we define Poisson brackets on the world sheet $\Sigma$.  We choose\cite{Arnlind:2012cx}
\be \{\xi_0, e^{i\xi^1}\}=\frac {ie^{i\xi^1}}{\sqrt{-{\tt g}}}\ee
More generally, the Poisson bracket of any two functions $f$ and $h$ of the string parameters $\xi$ is \be
\{f(\xi),h(\xi)\}=\frac 1{\sqrt{-{\tt g}}}\epsilon^{ab}\partial _a f\partial_b h\;\ee  With this definition the equations of motion (\ref{cleom}) can be re-expressed in terms of the Poisson algebra
\be \{\{x_\mu,x_\nu\},x^\nu\}+\frac 1{2\rho} \epsilon_{\mu\nu\rho}\{x^\nu,x^\rho\} =0 \label{clmeas}\ee 
The $dS^2$ solution (\ref{adstwo})  corresponds to the $SO(2,1)$ Poisson structure,
\be \{ x_\mu,x_\nu\} =\frac 1{2\rho} \epsilon_{\mu\nu\lambda} x^\lambda   \label{ds2pbs}\;, \ee 
and thus respect the $2+1$ Lorentz
symmetry of the background space.  On the other hand, for the static string solution (\ref{sss}), one gets the algebra of the two-dimensional Euclidean group
\be \{x^0,x^1\}=-\frac 1\rho x^2\qquad\quad \{x^0,x^2\}=\frac 1\rho x^1\qquad\quad \{x^1,x^2\}=0\label{sspbs}\ee This solution is invariant only under rotations in the plane and continuous time translations.

\section{Noncommutative static string}

\setcounter{equation}{0}

The equations of motion (\ref{clmeas}), and the solutions (\ref{ds2pbs}) and (\ref{sspbs}),  have a straightforward noncommutative generalization.   For this  we replace the three embedding coordinates $x^\mu$ by infinite-dimensional  self-adjoint matrices  $X^\mu$, and Poisson brackets by commutators.  Then the noncommutative version of  (\ref{clmeas}) is
\be [ [X_\mu,X_\nu],X^\nu]+i\alpha \epsilon_{\mu\nu\lambda}[X^\nu,X^\lambda] =0\label{eqofmot}\;,\ee where $\alpha$ is a constant with units of length.
(\ref{clmeas}) can be regarded as the commutative limit  of this matrix equation. For this we should introduce some noncommutativity parameter, say  $\theta$, such that $X^\mu\rightarrow x^\mu$ when  $\theta\rightarrow 0 $, and  to lowest order in   $\theta$,
$[f(X),h(X)]\rightarrow i\theta \{f(x),h(x)\}$.  In order to recover (\ref{clmeas}) we also need that $\alpha$  vanishes in the limit $\theta \rightarrow  0$, specifically, $\alpha\rightarrow \frac\theta{2\rho}$.
 The equations of motion (\ref{eqofmot}) are  invariant under: i) 
 Lorentz transformations $X^\mu\rightarrow \Lambda^\mu_{\;\;\nu} X^\nu$,  where $\Lambda$ is a $3\times 3$ Lorentz matrix,
 ii)  translations in the three-dimensional Minkowski space $X^\mu\rightarrow X^\mu+a^\mu\BI$, where $\BI$ is the  unit matrix, and  iii) unitary `gauge' transformations, $X^\mu\rightarrow UX^\mu U^\dagger$, where $U$ is an infinite dimensional unitary matrix.  In section 5, we include the fermionic sector so that the system has an additional ${\cal N}=1$ supersymmetry.

The equations of motion (\ref{eqofmot}) can be obtained from an action principle, with  the matrix model action given by\cite{Iso:2001mg}
\be S(X)=\frac 1{g^2}{\rm Tr}\Bigl(-\frac 14 [X_\mu, X_\nu] [X^\mu,X^\nu] +\frac 23 i \alpha \epsilon_{\mu\nu\lambda}X^\mu X^\nu X^\lambda\Bigr)\;\label{mmactn}\ee Here we introduce the coupling constant $g$  and
Tr is an invariant  trace.  The first term is the standard IKKT kinetic energy, while the second term is the matrix  analogue of (\ref{NScplng}).  Extremizing $S(X)$ with respect to variations in $X_\mu$ gives (\ref{eqofmot}).

The matrix analogue of the  $dS^2$ solution (\ref{ds2pbs}) is $X^\mu=X^\mu_{(dS)}$, where $X^\mu_{(dS)}$ are defined by the commutation relations 
 \be  [X^\mu_{(dS)},X^\nu_{(dS)}]= i \alpha \epsilon^{\mu\nu\lambda}X_{(dS)\lambda}  \; \label{ncds}\ee  The commutation relations are preserved under the action of the $2+1$ Lorentz group, and moreover, they define the $so(2,1)$ Lie algebra.   $X^\mu_{(dS)}X_{(dS)\mu}$ is a Casimir of the algebra, and so an irreducible representation is obtained by setting $X^\mu_{(dS)}X_{(dS)\mu}=4\rho^2\BI$. This solution has been studied previously in \cite{Ho:2000fy}  \cite{Jurman:2013ota}, while its Euclidean space counterpart, the fuzzy sphere, has been known for a long time.\cite{Madore:1991bw}-\cite{Iso:2001mg} 

The focus in this article is the matrix analogue of the static string solution (\ref{sspbs}).  It is given by $X^\mu=X^\mu_{(0)}$, where $X^\mu_{(0)}$ satisfy 
 \be  [X_{(0)0},X_{(0)\pm}]=\pm  2 \alpha X_{(0)\pm}  \qquad\qquad   [X_{(0)+},X_{(0)-}]=0\;, \label{ncbeta0}\ee and $X_{(0)\pm}=X_{(0)1} \pm i X_{(0)2}$.   This solution breaks the $2+1 $ Lorentz symmetry of the target space  to  $SO(2)\times \tilde T$. $SO(2)$ correspond to rotations in the plane, while $\tilde T$ refers to discrete time translations (whose action is defined  below).  The solution is known as the noncommutative or fuzzy  cylinder.  We note that this solution survives when the background metric is changed to a Euclidean metric.  The algebra generated by $X_{(0)\mu}$ is the two dimensional Euclidean group. $X_{(0)+}X_{(0)-}$ is in the center of the algebra, and in an  irreducible representation it  is proportional to the identity, 
\be X_{(0)+}X_{(0)-} =\rho^2 \BI\label{ncccsmr}\ee   $ \rho$ can now be regarded as the radius of the noncommutative cylinder, and we can write $X_{(0)+}=\rho e^{i\hat \phi}$, where $e^{i\hat \phi}$ is a unitary operator. Another central element is $\;\exp{\Bigl(\frac {\pi i}\alpha X_{(0)0}\Bigr) }$.  In an  irreducible  representation  it is a phase times the identity, $e^{i\phi_0}\BI$.  Irreducible representation are thus labeled by both $\rho$ and $\phi_0$, $\;\{\rho>0$,\, $0\le\phi_0<2\pi\}$.
The eigenvectors $v_n $  of  $X_{(0)0}$ satisfy
\be  X_{(0)0}\,v_n =\tau_n v_n\qquad\quad   X_{(0)\pm}\, v_n=\rho  v_{n\pm 1} \;,\ee where   $\tau_n=(2 n +\phi_0/\pi)\alpha $, $n$ running over all integers.   The eigenvectors $ v_n$ form  a basis for the irreducible representations of the noncommutative cylinder.  Here one sees that the action of the discrete time translation operator $\tilde T$ is just  $\tau_n\rightarrow \tau_n +2\alpha \Delta n$, where $\Delta n$ is an integer.

\section{Perturbations about the  noncommutative cylinder}

\setcounter{equation}{0}

 Next we consider perturbations about the  noncommutative cylinder,
\be X_\mu=X_{(0)\mu}+2\alpha  A_\mu\;,\label{prtrbncc}\ee
 where $ A_\mu$ are fields on the noncommutative cylinder in some irreducible representation.   $ A_\mu$ are thus functions of $e^{i\hat \phi}$ and $X_{(0)0}$.  It is useful to define 
 the  field strengths 
 \beqa  F_{+-}&=&\frac 1{4{\alpha}^2}[X_+,X_-] \cr&&\cr
 F_{\pm  0}&=&\frac 1{4{\alpha}^2}[X_\pm,X_0]\pm\frac 1{2\alpha}X_\pm\label{fldstrngth}
\eeqa
 where   $X_\pm=X_1\pm iX_2$. 
  They transform covariantly under unitary gauge transformations and vanish when $A_\mu=0$.  $ A_\mu$ and  $ F_{\mu\nu}$ are defined to be dimensionless.  Upon substituting (\ref{prtrbncc}) and (\ref{fldstrngth}) into the action (\ref{mmactn}), one gets
\beqa S(X)&=&\frac {16\alpha^4}{g^2}{\rm Tr}\Bigl\{ \frac 18(F_{+-})^2+\frac 12F_{+0}F_{-0}
-\frac 12\Bigl(
A_+F_{-0}-A_-F_{+0}-A_0F_{+-}\Bigr)\cr &&\cr &&\quad\qquad -\frac 1{4\alpha}\Bigl(
X_{(0)+}[A_-,A_0]-X_{(0)-}[A_+,A_0]-X_{(0)0}[A_+,A_-]\Bigr)\cr &&\cr &&\quad\qquad -\frac 12 A_+A_-\Bigr\}\; +\; S(X_{(0)})\label{111}
\eeqa
The action   evaluated for the solution, $S(X_{(0)})$, is singular.

We  next replace the potentials and field strengths by their corresponding  Weyl-symbols, and their matrix products by the Moyal-Weyl star product on the cylinder. The latter was given in \cite{Chaichian:2000ia}, and is reviewed in appendix A.  So now $A_\mu$ denote   functions of a phase $e^{i\phi}$ and $\tau_n$, and the Weyl symbols of the field strengths (\ref{fldstrngth}) are given by
 \beqa  F_{+-}&=&\frac \rho{2{\alpha}}\Bigl(    [e^{i\phi},A_-]_\star-[e^{-i\phi},A_+]_\star\Bigr) +[A_+,A_-]_\star
 \cr&&\cr
 F_{\pm 0}&=&\frac \rho{2{\alpha}}   [e^{\pm i\phi},A_0]_\star+[A_\pm ,A_0]_\star +(i\partial_\phi\pm 1)A_\pm
\;,\label{ncfldstr}
\eeqa
 where $[\;,\;]_\star$ denotes the star commutator.  
In obtaining  the action we replace the trace in (\ref{111}) by 
$\frac 1{2\pi}  \int_{-\pi}^\pi d\phi \,\sum _{n=0,\pm1,\pm2,...}$, where the sum is over all eigenvalues $\tau_n$ of  $X_{(0)0} $.  Thus
\beqa S(X)&=&\frac {8\alpha^4}{\pi g^2} \int_{-\pi}^\pi d\phi \,\sum _{n=0,\pm1,\pm2,...}\Bigl\{ \frac 18(F_{+-})_\star^2+\frac 12F_{+0}\star F_{-0}\cr &&\cr &&\qquad\qquad\qquad
-\frac 12\Bigl(
A_+\star F_{-0}-A_-\star F_{+0}-A_0\star F_{+-}\Bigr)\cr &&\cr &&\qquad\qquad\qquad -\frac \rho{4\alpha}\Bigl(
[ e^{i\phi},A_-]_\star - [e^{-i\phi},A_+]_\star  \Bigr)A_0-\frac i2 \partial_\phi A_+ \star A_-\;-\frac 12 A_+\star A_-\Bigr\}\cr &&\cr &&\qquad \qquad\qquad+ S(X_{(0)})\label{117}\;,
\eeqa
where $F_\star^2=F\star F$.
The resulting equations of motion are
\beqa  D_+F_{-0}+  D_-F_{+0} &=&-iF_{+-}\cr &&\cr
D_+F_{+-}-2D_0 F_{+0}&=&0
\;,\eeqa 
where  the noncommutative covariant derivatives are defined by
\beqa  D_0 F &=& \partial_\phi F +i[A_0,F]_\star\cr &&\cr
 D_\pm F &=& i\Bigl[\frac\rho{2\alpha} e^{\pm i\phi} + A_\pm,F\Bigr]_\star\;, \label{nccvrntdrv}\eeqa

The dynamical degrees of freedom $A_\mu$, $\mu=0,1,2$, are noncommutative gauge potentials in the $2+1$ dimensional target space.  The system can also be expressed in terms of  gauge potentials on
the noncommutative cylinder.  Of course, there are only two of the latter, which we denote by $a_\tau$ and $a_\phi$.  This means that $A_\mu$ contains one additional degree of freedom, which we call $b$ and assume to be in the adjoint representation of the noncommuative gauge group.  In order to make the identification, we compare noncommutative gauge transformations of $A_\mu$  with those of  $a_\tau$,  $a_\phi$ and $b$.  Infinitesimal gauge variations of the former have the form $\delta A_\mu =D_\mu\lambda$, $\lambda$ being an  infinitesimal function of $\tau_n$ and $e^{i\phi}$. Consequently, $\delta F_{\mu\nu}=- i[\lambda,F_{\mu\nu}]_\star \,$. 
For the latter we want 
 \beqa \delta a_\phi &=&\partial_\phi\lambda +i[a_\phi,\lambda]_\star\cr&&\cr
\delta e^{i(\phi-2\alpha a_\tau)} &=&i[e^{i(\phi-2\alpha a_\tau)},\lambda]_\star\cr&&\cr
\delta b &=&i[b,\lambda]_\star\label{nctpofab}
\eeqa
It is evident that $A_0$ transforms as $a_\phi$, and so we  identify the two, $A_0=a_\phi$.  Therefore
$A_\pm$  contain both $a_\tau$ and $b$.  An identification  which is consistent with the variations (\ref{nctpofab}) is
\beqa   
 2\alpha A_+&=& (\rho+2\alpha b)\star e^{i(\phi-2\alpha a_\tau)}-\rho e^{i\phi} 
\cr&&\cr
 2\alpha A_-&=& e^{-i(\phi-2\alpha a_\tau)}\star (\rho+2\alpha b)-\rho e^{-i\phi}\;,\label{ApAm}
\eeqa
where $\rho e^{\pm i\phi}$ was subtracted off on the right hand side in order to have the correct  commutative limit $\alpha\rightarrow 0$.

We now  examine the action in the  commutative  limit $\alpha\rightarrow 0$.
The eigenvalues $\tau_n$ become continuous in this limit, $\tau_n\rightarrow\tau$, and so we recover the commutative cylinder, while the sum  in (\ref{117}) is replaced by
$\frac 1{2\alpha}\int d\tau$.  The potentials  $ (a_{\tau},a_{\phi})$ on the cylinder  undergo standard $U(1)$ gauge variations, $ \delta (a_{\tau},a_{\phi})=(\partial_{\tau}\lambda,\partial_{\phi}\lambda)$ in the limit,
and $b$ is gauge invariant at zeroth order in $\alpha$.  From (\ref{ApAm}) we get 
\be A_\pm\rightarrow(  b \mp i \rho a_\tau ) e^{\pm i\phi}\label{BgAintltla}\;,\ee in addition to $ A_0= a_\phi$.  
Substituting in (\ref{ncfldstr}) and taking the commutative limit gives \be F_{+-}\rightarrow -2\rho\partial_\tau b\qquad\quad   F_{\pm 0}\rightarrow (  i \partial_\phi b\mp \rho f_{\tau\phi} ) e^{\pm i\phi}\;,\label{BgFintltlf}\ee
 where $ f_{\tau\phi}=\partial_\tau a_\phi- \partial_\phi a_\tau $ is the $U(1)$ field strength on the cylinder.
Finally, substituting (\ref{BgAintltla}) and (\ref{BgFintltlf}) into  the action (\ref{117}) and taking the commutative limit gives
\be S(X)- S(X_{(0)})\rightarrow \frac {4\alpha^3}{\pi g^2} \int_{-\pi}^\pi d\phi \int^\infty_{-\infty}d\tau\, \Bigl(-\frac{\rho^2}2(f_{\tau\phi})^2+\frac {\rho^2}2 (\partial_\tau b)^2-\frac 1{2} (\partial_\phi b)^2-\rho \,bf_{\tau\phi} \Bigr)\;\label{121}
\ee 
It describes a  scalar field coupled to a $U(1)$ gauge field on the cylinder.  In comparing with the corresponding noncommutative limit for the fuzzy sphere,\cite{Iso:2001mg} it is interesting to not that, unlike there, no explicit mass term appears for the scalar field.
 However, the $U(1)$  gauge field is not dynamical in one spatial dimension, and can be eliminated using the equations of motion,  
\be f_{\tau\phi}=-\frac b\rho +{\rm constant}  \ee
The result is a tachyonic mass term for the scalar field.  The tachyonic mass is $m^2=-1/\rho^2$, which  vanishes in the infinite radius limit.

 Nonlocal interactions will appear upon including noncommutative corrections to the action (\ref{121}), but they are  unlikely to cure the tachyonic instability present at lowest order.   Actually, the presence of a tachyon may be traced to the instability of the classical solution (\ref{sss}).  The instability arises from uniform adiabatic excitations of the radial degree of freedom.  For this, replace the constant $\rho$ in (\ref{sss}) by $\rho+ 2\alpha b(\xi^0,\xi^1)$. One can compute the energy for this string configuration using the previous definition (\ref{strsnrg})  for the stress-energy tensor, where we  identify $\xi^0=\tau$ and $\xi^1=\phi$.  After expanding in $\alpha$, one gets \be\int dy^1dy^2\, T^{00}(y)=\pi \rho {\cal T}\biggl\{1+ 4\alpha^2\Bigl((\partial_0 b)^2 +\frac 1{\rho^2}(\partial_1 b)^2- \frac 1{\rho^2}b^2\Bigr)+{\cal O}(\alpha^3)\biggr\}\ee
Thus, since this expression is not positive definite, the static closed  string is unstable, specifically due to uniform adiabatic deformations in the radius. In section 6, we shall introduce a fourth order  term to the matrix model which explicitly breaks the $SO(2,1)$  symmetry and eliminates the tachyonic mass.

\section{Nonabelian and supersymmetric extensions}

\setcounter{equation}{0}

Before discussing the cure to the above instability, we give two standard extensions of the model.  In the first, we generalize  to  a stack of $N$ coinciding branes,  and in the second, we  include the fermionic sector in an ${\cal N}=1$ supersymmetric theory. 
\subsection{Generalization to $U(N)$ }
The usual  generalization to $U(N)$ gauge theory is possible.  For this one examines a solution to (\ref{eqofmot}) 
of the form \be
X_\mu=X_{(0)\mu} \otimes\BI_N\;,\ee
$\BI_N$ being the identity matrix in $N$ dimensions and $X_{(0)\mu}$ defined by (\ref{ncbeta0}) and (\ref{ncccsmr}).
  This solution is associated with a stack of $N$ coinciding branes.  General  perturbations about the solution are of the form  $ X_\mu=X_{(0)\mu} \otimes\BI_N+2\alpha  A_\mu^{{\tt a}}\otimes T_{\tt a}$, $\,{\tt a}=1,2,3,...N^2,\,$ where
 $T_{\tt a}$ are $N\times N$ hermitean matrices generating  $U(N)$.  We will assume they are normalized according to Tr$\;T_{\tt a} T_{\tt b} =\delta_{\tt a, b}$.  The nonabelian generalization of the action (\ref{117}) on the noncommutative cylinder is straightforward.  One can  re-express $A_\mu^{{\tt a}}$ in terms of  nonabelian gauge potentials ($a_\tau^{\tt a},a_\phi^{\tt a}$) on the noncommutative cylinder, along with additional fields $ b^{\tt a}$ which transform in the adjoint representation of the noncommutative gauge group.
In the commutative  limit $\alpha\rightarrow 0$, the identification is
\be A_\pm\rightarrow( b^{\tt a} \mp i \rho a_\tau^{\tt a} ) e^{\pm i\phi}\otimes T_{\tt a}\;,\ee in addition to $ A_3= a_\phi^{\tt a}\otimes T_{\tt a} $.  This is the generalization of  (\ref{BgAintltla}).
Then 
\be F_{+-}\rightarrow -2\rho({\tt D}_\tau b)^{\tt a}\otimes T_{\tt a}\qquad\quad   F_{\pm 3}\rightarrow\Bigl( ( {\tt D}_\phi b)^{\tt a}\mp \rho f_{\tau\phi}^{\tt a} \Bigr) e^{\pm i\phi}\otimes T_{\tt a} \;,\ee where the Yang-Mills fields and the covariant exterior derivative of the scalar  are defined respectively by  $f_{\tau\phi}^{\tt a}=\partial_\tau a_\phi^{\tt a}- \partial_\phi a_\tau ^{\tt a}- {\tt C}^{\tt a}_{\tt b c} a^{\tt b}_\tau a_\phi^{\tt c}\;$ and
 $\;({\tt D} b)^{\tt a}=d b^{\tt a}-  {\tt C}^{\tt a}_{\tt bc} a^{\tt b} b^{\tt c}$. Here  $  {\tt C}^{\tt a}_{\tt bc}  $ are the $u(N)$ structure constants, $[T_{\tt b},T_{\tt c}]=i  {\tt C}^{\tt a}_{\tt bc}  T_{\tt a}.\,$ Then the action on the cylinder  (\ref{121})  has the obvious  $U(N)$  generalization
\beqa S(X)- S(X_{(0)})&\rightarrow&\frac {4\alpha^3}{\pi g^2} \int_{-\pi}^\pi d\phi \int^\infty_{-\infty}d\tau\, \Bigl(-\frac{\rho^2}2 f_{\tau\phi}^{\tt a} f_{\tau\phi}^{\tt a}+\frac{\rho^2}2 ({\tt D}_\tau b)^{\tt a} ({\tt D}_\tau b)^{\tt a}-\frac 1{2} ({\tt D}_\phi b)^{\tt a} ({\tt D}_\phi b)^{\tt a}\cr &&\cr
&&\qquad\qquad\qquad\qquad\qquad -\rho\, b^{\tt a} f_{\tau\phi}^{\tt a}
 \Bigr)
\eeqa Again,  the gauge fields are not dynamical in one spatial dimension.  The  field equation for  $ f^{\tt a}_{\tau\phi}$ states that   $ f^{\tt a}_{\tau\phi}+\frac 1\rho b^{\tt a}$ is covariantly  constant, while the  equations for $b^{\tt a}$ yield $N^2$ tachyons.

\subsection{Fermionic sector}

We now consider fermionic degrees of freedom in the  ${\cal N}=1$ suspersymmetric extension of the matrix model.  They are denoted by the infinite dimensional hermitean matrix $\Psi$, whose matrix elements $\Psi_{AB},\; A,B=1,2,...$ are complex Majorana spinors in $2+1$ Minkowski space.  More precisely, $\Psi_{AB}=\Psi^*_{BA}=\xi_{AB}^1+i\xi_{AB}^2$, where $\xi_{AB}^1$ and $\xi_{AB}^2$ are real Majorana spinors in $2+1$ Minkowski space.  The spinors are acted on by a  real   $2\times 2$ representation  of the
$\gamma$-matrices,  satisfying $\{\gamma^\mu,\gamma^\nu\}= 2\eta^{\mu\nu}\BI_2$.  Our conventions for the spinors and gamma matrices are those of \cite{RuizRuiz:1996mm}, which are reviewed in  appendix B.  As usual,
$\Psi$ transforms in the adjoint representation of the infinite dimensional unitary  group.  We  take the components of $\Psi$   to have units of length to the three-halves power.  The addition of the terms
\be S_F(X,\Psi)=\frac 1{g^2}{\rm Tr}\Bigl( \frac 12\bar\Psi\gamma^\mu[X_\mu,\Psi] +i\alpha\bar\Psi\Psi\Bigr)\;,\label{fcmmactn}\ee 
 to the bosonic action (\ref{mmactn}) is consistent with ${\cal N}=1$ supersymmetry, where infinitesimal variations are given by 
\be \delta X_\mu=\bar \epsilon \gamma_\mu\Psi\qquad\qquad \delta \Psi
=-\frac 12 [X_\mu,X_\nu]\gamma^{\mu}\gamma^\nu\epsilon \;\ee  Here $\epsilon $ is an infinitesimal real Majorana spinor.  To verify this, the supersymmetry variation of the first term in the trace in (\ref{fcmmactn}) gives
\be  \frac 12\delta\Bigl( {\rm Tr}\bar\Psi\gamma^\mu[X_\mu,\Psi]\Bigr)= {\rm Tr}\bar\Psi[X^\mu, [X_\mu,X_\nu]]\gamma^\nu\epsilon\;,\ee
after using the identities (\ref{btwo}) and  (\ref{bfour}) in appendix B.  This result is canceled by the corresponding supersymmetric variation of the quartic term in the trace in (\ref{mmactn}), i.e., $-\frac 14{\rm Tr} [X_\mu, X_\nu]^2\,$.  Similarly, the supersymmetric variation of the `mass' term in  (\ref{fcmmactn}) gives
\be  \delta\, {\rm Tr}\bar\Psi\Psi=-\epsilon^{\mu\nu\lambda}\, {\rm Tr}\bar\epsilon\gamma_\mu\Psi[X_\nu,X_\lambda]\;,\ee  using  (\ref{btwo}) and (\ref{bthree}), which  is canceled by the corresponding supersymmetric variation of the cubic term in the trace in (\ref{mmactn}), i.e., $\frac 23  \epsilon_{\mu\nu\lambda}{\rm Tr}X^\mu X^\nu X^\lambda\,$.

The fermionic contributions (\ref{fcmmactn}) contribution to the action introduce a source term to the right hand side of the equation of motion (\ref{eqofmot}).  The  fuzzy deSitter
solution 
(\ref{ncds}) and noncommutative cylinder solution (\ref{ncbeta0}) remain valid when $\Psi=0$. 

Next we examine perturbations  (\ref{prtrbncc}) about the  noncommutative cylinder solution with $\Psi=0$.  For the perturbations, it is convenient to rescale the fermion field, using $\Psi=\frac1 4{\sqrt{\rho\alpha^{3}}}\,\psi, $ where $\psi$ is a function on the noncommutative cylinder in some irreducible representation.  Then we  can write the fermionic addition to the action (\ref{117}) in terms of symbols of the fields on the noncommutative cylinder.  We get
 \beqa S_F(X,\Psi)&=&\frac {8\alpha^4}{\pi \rho g^2} \int_{-\pi}^\pi d\phi \sum _{n=0,\pm1,\pm2,...}\Bigl\{ -i\bar\psi\star \Bigl(\gamma^0 D_0\psi+\frac 12\gamma^-D_+\psi+\frac 12\gamma^+ D_-\psi\Bigr) +i\bar\psi\star \psi\Bigr\}\;,\cr &&\label{538}\eeqa 
where $\gamma^\pm=\gamma^1\pm i\gamma^2$ and the covariant derivatives are given in (\ref{nccvrntdrv}).
In the commutative  limit $\alpha\rightarrow 0$, the covariant derivatives reduce to ordinary derivatives, $ D_0\psi \rightarrow \partial_\phi\psi$ and $ D_\pm\psi \rightarrow \mp i\rho e^{\pm i\phi} \partial_\tau\psi$, and the fermionic action (\ref{538})  reduces to
\beqa S_F(X,\Psi)&\rightarrow &\frac {4\alpha^3}{\pi g^2} \int_{-\pi}^\pi d\phi \int^\infty_{-\infty} d\tau\;\Bigl\{ \bar\psi\Bigl(\tilde\gamma^\tau \partial_\tau\psi+\frac 1{\rho}\tilde \gamma^\phi \partial_\phi\psi\Bigr) +\frac i{\rho}\bar\psi\psi\Bigr\}\;,\eeqa
where the gamma matrices on the cylinder $(\tilde \gamma^\tau,\tilde \gamma^\phi)$ are defined by $\tilde \gamma^\tau =i\gamma^2\cos\phi  -i\gamma^1\sin\phi  $ and  $\tilde \gamma^\phi =-i \gamma^0$.  The  action
leads to the Dirac equation
\be\tilde\gamma^\tau \partial_\tau\psi+\frac 1{\rho}\tilde \gamma^\phi \partial_\phi\psi +\frac i{\rho}\psi=0
\ee
on the cylinder.
Using the conventions in the appendix, it can be checked that the fermionic degrees of freedom, like the bosonic degree of freedom, are tachyonic.  Also, like with the bosonic field, the tachyonic mass is $m^2=-1/\rho^2$, which  vanishes in the infinite radius limit.

\section{Explicit symmetry breaking}

\setcounter{equation}{0}

Here we show how one can remove tachyonic  mass term that appeared in the commutative limit of the model in section 4.  Our proposal is to   add  an explicit symmetry breaking term to the bosonic sector of the matrix model action  (\ref{mmactn}).  The term breaks Lorentz symmetry, as well as supersymmetry, but is consistent with the invariance group of the  noncommuative cylinder solution, i.e., $SO(2)\times \tilde T$.  One possibility is that we add  the quadratic term Tr$X_+X_-$ to  (\ref{mmactn}).  It can be checked that there can be up  to two noncommutative cylinder solutions to the equations of motion following from this modification of the action.  However, after perturbing around these solutions and taking the commutative limit one finds that the mass of the scalar field $b$   is not modified.  In order to modify the mass we instead need to add  a higher order term to the matrix action.  Below we examine a quartic symmetry breaking term.  The total action is
\beqa S_{sb}(X)&=&S(X)+\frac{ \beta}{4g^2}{\rm Tr}(X_+X_-)^2\cr & &\cr &=&\frac 1{g^2}{\rm Tr}\Bigl(\frac 18 [X_+,X_-]^2+\frac 12 [X_+,X_0][X_-,X_0]  + \tilde\alpha X_0[X_+,X_-]+\frac{\beta}4(X_+X_-)^2
\Bigr)
\;,\label{mmactnweb}\cr&&\eeqa 
where  $\beta$ is a real constant and we now denote the coefficient of the cubic term by $\tilde\alpha$ in order to distinguish it from the noncommutative parameter $\alpha$ appearing in the classical solution (\ref{ncbeta0}).  The $2+1$ Lorentz symmetry of the background space is reduced  to rotational  symmetry on the plane due to the last term. (If one includes the fermionic contribution (5.5) to the action, then supersymmetry is broken as well.) We will see that  $1+1$ space-time symmetry of the  cylinder is preserved in the commutative limit.  

The   equations of motion  now read
\beqa 
[[X_+,X_0],X_-]+[[X_-,X_0],X_+]&=& 2\tilde \alpha [X_-,X_+] \cr &&\cr
 [[X_0, X_+], X_0] +\frac 12 [[X_+,X_-],X_+] &=& 2\tilde\alpha [X_+,X_0] -\beta X_+X_- X_+
\eeqa  They reduce to (\ref{eqofmot}) upon setting $\beta=0$ and $\tilde\alpha=\alpha$.
The noncommutative deSitter configuration (\ref{ncds}) does not satisfy the equations of motion when $\beta\ne 0$.  On the other hand, there exist two noncommutative cylinder solutions of the form (\ref{ncbeta0}), provided that $\tilde\alpha^2+\beta\rho^2>0$  (and one when $\tilde\alpha^2+\beta\rho^2=0$).  $\rho$ again denotes the radius  of the cylinder. The noncommutative parameters $\alpha=\alpha_\pm$ for the two solutions are given by
\be \alpha_\pm=\frac 1 2(\tilde\alpha\pm \sqrt{\tilde\alpha^2+\beta\rho^2} \,) \label{2solns}\ee
We can consider various limits. The two solutions reduce to the  solution of the previous section ($\alpha_+\rightarrow\tilde\alpha$) and the vacuum solution ($\alpha_-\rightarrow 0$)  in the limit $\beta\rightarrow 0$.   The solutions are degenerate, $\alpha_\pm\rightarrow \tilde\alpha/2$, in the limit $\tilde\alpha^2+\beta\rho^2\rightarrow 0$.

 Perturbations (\ref{prtrbncc}) can be considered about either of the two noncommutative cylinder solutions, associated with some irreducible representation.  The latter are again labeled by the radius $\rho$ and and phase angle $\phi_0$.  Upon substituting into the action  (\ref{mmactnweb}) and using the previous definitions (\ref{ncfldstr}) for the noncommutative field strengths, we get
\beqa S_{sb}(X)&=&\frac {8\alpha^4}{\pi g^2} \int_{-\pi}^\pi d\phi \,\sum _{n=0,\pm1,\pm2,...}\biggl\{ \frac 18(F_{+-})_\star ^2+\frac 12F_{+0}\star F_{-0}\cr &&\cr &&\qquad \qquad\qquad \qquad\qquad  
-\frac 12\Bigl(
A_+\star F_{-0}-A_-\star F_{+0}-\frac {\tilde\alpha}\alpha A_0\star F_{+-}\Bigr)\cr &&\cr &&\qquad\qquad \qquad\qquad\qquad -\frac \rho {4\alpha}\Bigl(
 e^{i\phi}[A_-,A_0]_\star-e^{-i\phi}[A_+,A_0]_\star\Bigr)-\frac {i\tilde\alpha}{2\alpha}\partial_\phi A_+\star A_-\;\cr &&\cr &&\qquad \qquad\qquad \qquad\qquad+\frac \beta{16\alpha^2}\Bigl(\rho(A_+\star e^{-i\phi}+A_-\star e^{i\phi})+2\alpha A_+\star A_-\Bigr)_\star^2\cr &&\cr &&\qquad \qquad\qquad \qquad\qquad  +\frac 12\Bigl(\frac{\beta\rho^2}{4\alpha^2}-1\Bigr) A_+\star A_-\biggr\}\;+\; S_{sb}(X_{(0)})\;,
\eeqa
where $\alpha$ stands for either  $\alpha_+$ or $\alpha_-$, depending  upon which  solution is expanded around.  This action reduces to (\ref{117}) when $\beta= 0$ and $\tilde\alpha= \alpha$.  
In section 4, the commutative limit  corresponded to  $\alpha\rightarrow 0$. Now since there are two noncommutative cylinder solutions, we can have $\alpha_-\rightarrow 0$,   $\alpha_+\rightarrow 0$ or both.  In the generic case where both go to zero, this means that the constants $\tilde \alpha$ and  $\sqrt{\beta}$ are of order $\alpha\sim\alpha_\pm$, using (\ref{2solns}).\footnote{Another possibility is that we keep $\tilde\alpha$ finite, while $\beta$ tends to zero.  Then $\alpha_-\rightarrow -\rho^2\tilde\alpha\beta/4$, while $\alpha_+$ remains finite.}  The limiting values are constrained by the condition  $\frac{\beta\rho^2}{4\alpha^2}-1+\frac {\tilde\alpha}\alpha =0$.
We assume that at least one cylindrical solution  (\ref{2solns}) exists when taking the limit, i.e., that $\alpha$ is real. Taking $\alpha,\;\beta$ and $\tilde\alpha\;\rightarrow 0$, while keeping $\tilde\alpha /\alpha$ and $\beta /\alpha^2$ finite, yields
\beqa S_{sb}(X)- S_{sb}(X_{(0)})&=&\frac {4\alpha^3}{\pi g^2} \int_{-\pi}^\pi d\phi \int^\infty_{-\infty} d\tau\,\biggl\{\frac {\rho^2}2(\partial_\tau b)^2-\frac 12 (\partial_\phi b)^2-\frac {\rho^2}2(f_{\tau\phi})^2
\cr &&\cr &&\qquad\qquad\qquad\quad\qquad -\Bigl(1+\frac{\beta\rho^2}{4\alpha^2}\Bigr)\rho b f_{\tau\phi}
+\frac {\beta\rho^2}{4\alpha^2}b ^2\;\biggr\}\;\label{647}
\eeqa
  The last term in the integral gives a new contribution to the mass-squared of the scalar field.  Since the sign of $\beta$ is not restricted, neither is the sign of this contribution.  To determine the total effective  mass of $b$, we should again eliminate the nondynamical gauge field using its equations of motion. Here we get 
\be f_{\tau\phi}=-\Bigl(1+\frac{\beta\rho^2}{4\alpha^2}\Bigr)\frac b\rho +{\rm constant}  \;,
\ee
which can be substituted back into the action  (\ref{647}).  The last three terms in the integrand combine to give a mass  term for the scalar, $-\frac {m^2}2 b^2$, where
\be 
\rho^2 m^2=-1-\frac{\beta\rho^2}{\alpha^2}-\frac 1{16}\Bigl(\frac{\beta\rho^2}{\alpha^2}\Bigr)^2 \ee
Zero total mass occurs for $\frac {\tilde \alpha}\alpha=  3\mp\sqrt{3}$, or equivalently, 
 $\frac{\beta\rho^2}{4\alpha^2}=-2\pm\sqrt{3}$.  More generally, the tachyonic mass is eliminated for the following limiting values for $\frac {\tilde \alpha}\alpha$ and  $\frac {\beta\rho^2}{4\alpha^2}$:
\be  3-\sqrt{3}\le\frac {\tilde \alpha}\alpha\le  3+\sqrt{3}\qquad\quad  -2-\sqrt{3}\le \frac{\beta\rho^2}{4\alpha^2}\le-2+\sqrt{3}\ee The maximum value for the mass-squared in this model is $3/\rho^2$, which occurs for $\frac {\tilde \alpha}\alpha\rightarrow 3$, or equivalently $\frac {\beta\rho^2}{4\alpha^2}\rightarrow -2$. 

Concerning the fermionic sector, one can, of course,  eliminate the tachyonic mass for the fermion of the previous section
by modifying the second term in the trace in (\ref{fcmmactn}), again violating supersymmetry.

\section{A BF matrix model}
In the previous section we allowed for the addition of  a symmetry breaking term in the action.
An interesting limit of this system corresponds to the kinetic energy term, i.e, the first term in the trace of (\ref{mmactn}),  vanishing.  For the example of  a quadratic symmetry breaking term [instead of the quartic term in (\ref{mmactnweb})], the action  would reduce to
\be S_{BF}(X)=\frac 1{g^2}{\rm Tr}\Bigl(\tilde\alpha X_0[X_+,X_-]+\frac{\tilde\beta}2X_+X_-
\Bigr)
\;\label{tpmmatn}\ee 
 in the  limit of zero kinetic energy.  $\tilde\beta$ is  the coefficient of the symmetry breaking term.  The resulting matrix model is somewhat analogous to that considered in \cite{Chaney:2013vha}, which was utilized for the purpose of obtaining the BTZ black hole entropy.  The major difference between the two systems is  that  time is a continuous parameter in the previous model, which thus describes a zero-brane.

  The equations of motion following from (\ref{tpmmatn}) have a noncommutative cylinder solution (\ref{ncbeta0}), where  the noncommutative parameter is given  by $\alpha=-\frac{\tilde\beta}{4\tilde\alpha}$.  (As in the previous section, the symmetry breaking term does not allow for the fuzzy de Sitter solution.)  As before,  $\rho$ and $\phi_0$ label the irreducible representations  of the solutions, and    the spectrum of  $X_{(0)0}$ is given by $\{\tau_n=(2 n +\phi_0/\pi)\alpha,\,\; n=0,\pm1,\pm2,...\}$.  Perturbations about this solution gives
\be S_{BF}(X)- S_{BF}(X_{(0)})=\frac {4\alpha^3{\tilde\alpha}}{\pi g^2} \int_{-\pi}^\pi d\phi \,\sum _{n=0,\pm1,\pm2,...}\biggl\{
  A_0\star  F_{+-}
- {i  }\partial_\phi A_+\star A_- +\frac{\tilde\beta}{4\alpha\tilde\alpha} A_+\star A_-\biggr\}\;,
\ee
where $A_\mu$ are again the symbols for the noncommutative potentials in $2+1$ target space.  The field strengths $F_{\pm 0}$ do not appear in the action.
In the noncommutative limit,  $\alpha=-\frac{\tilde\beta}{4\tilde\alpha}\rightarrow 0$, one gets 
\be S_{BF}(X)-S_{BF}(X_{(0)})\rightarrow -\frac {4\alpha^3c}{\pi g^2} \int_{-\pi}^\pi d\phi \int^\infty_{-\infty}d\tau\, \;
\rho \, bf_{\tau\phi}\;,
\ee where $f_{\tau\phi}$  and $b$ are again the gauge  and scalar field, respectively, with the identifications (\ref{ApAm}) and  
$c=4\tilde\alpha^2/\tilde\beta$.  Not surprisingly, the kinetic energies for the gauge  and scalar field are absent, and all that remains is a $BF$ action on the cylinder.  The resulting system has a  global degree of freedom, $\int_{-\pi}^\pi d\phi \,a_\phi $, which can be interpreted as the magnetic flux through the cylinder.

\section{Concluding remarks}

\setcounter{equation}{0}

We have shown that   perturbations around the noncommuative cylinder  in the matrix action (\ref{mmactn}) lead to a tachyonic scalar field. This was attributed to the instability of the circular static string  under uniform adiabatic deformations.
In section 6 we introduced a stabilizing term in the action which removes the the tachyonic mass term for the scalar field.  In these discussions we focused our attention predominantly on the commutative limit.  The full noncommutative theory has  nonlocal interactions.  The effective theory for the scalar field is highly nontrivial since it also involves couplings to the nondynamical gauge fields.  Field theories on the noncommutative cylinder have been investigated previously, and have been shown to exhibit issues with  unitarity.\cite{Chaichian:2000ia}  It is of interest to investigate whether such problems can be resolved in the context of the matrix model discussed here. 

There are various possible generalizations of the matrix action (\ref{mmactn}),  including the cubic term,  to higher space-time dimensions.  One  is that  $\epsilon_{\mu\nu\lambda}$ in (\ref{eqofmot}) is replaced by other Lie algebra structure constants.  For the case of the $su(n)$ algebra, $\mu,\nu,...$ take on $n^2-1$ values, leading to an $n^2-1$ dimensional target space.  When the signature is Euclidean, the resulting theory has  fuzzy  $CP^{n-1}$ solutions, which preserve the full $SU(n)$ symmetry of the target space.\cite{Balachandran:2005ew} Other solutions, which break the symmetry and thus are analogous to the noncommutative cylinder, should exist as well.  
 Another possible generalization is that we rewrite the action in terms of the noncommutative analogue of antisymmetric, or Neveu-Schwarz fields $\;B_{\mu\nu}$, which here are infinite dimensional hermitean matrices.  In three space-time dimensions, we can take them to be just the space-time dual of the hermitean matrices $X^\mu$,   $\;B_{\mu\nu}=\epsilon_{\mu\nu\rho} X^\rho$.  The action (\ref{mmactn}) in terms of these matrices reads
\be S(B)=\frac 1{g^2}{\rm Tr}\Bigl(-\frac 1{16} [B_{\mu\nu},B_{\rho\sigma}]^2-\frac 23 i \alpha\; B_{\mu\nu} B^{\nu\rho}B_\rho^{\;\;\mu}\Bigr)\ee
Expressed in this manner, the matrix action easily generalizes to any number of space-time dimensions. In $d+1 $ space-time there are a total of $d(d+1)/2$ hermitean matrices $B_{\mu\nu}$.  Solutions,  such as  noncommutative de Sitter space and tensor products of  noncommutative cylinders, fuzzy spheres and de Sitter spaces, appear in this model.   Fuzzy $CP^{n-1}$ solutions\cite{Balachandran:2005ew}  may result from this model as well.      We plan to pursue such generalizations in a later article.

\bigskip
\appendix
\appendice{\Large{\bf $\quad$   Star product on the noncommutative cylinder}}

Following \cite{Chaichian:2000ia},  one can define the analogue of a Moyal-Weyl star product  for  fields $\Phi(e^{i\hat \phi},X_{(0)0})$ on the noncommutative cylinder in any given irreducible representation.  For this, 
 expand  $ \Phi$ in terms of the noncommutative analogue of Fourier modes \be e^{i(k\hat \phi-\omega X_{(0)0})},\quad k=0,\pm1,\pm2,...\;,\quad -\frac\pi{2\alpha}<\omega\le\frac\pi{2\alpha} \;,\ee
subject to the orthonormality conditions
\be\frac  \alpha{ \pi}\,{\rm Tr}\;e^{i(-k'\hat \phi+\omega' X_{(0)0})}\,e^{i(k\hat \phi-\omega X_{(0)0})}= \delta_{k,k'}\delta({\omega'-\omega})\label{orthonorm}\ee 
 Thus
\be  \Phi(e^{i\hat \phi},X_{(0)0})=\sum_{k=0,\pm1,\pm2,...}\int_{-\frac\pi{2\alpha}}^{\frac\pi{2\alpha}}d\omega\,\tilde \Phi_k(\omega)e^{i(k\hat \phi-\omega X_{(0)0})}\ee
The Fourier coefficients
$\tilde\Phi_k(\omega)$ are used to define the analogue of  Weyl symbols $ \Phi(e^{i \phi},\tau_n)$ on the cylinder \be  \Phi(e^{i \phi},\tau_n)=\sum_{k=0,\pm1,\pm2,...}\int_{-\frac\pi{2\alpha}}^{\frac\pi{2\alpha}}d\omega\,\tilde \Phi_k(\omega)e^{i(k\phi-\omega \tau_n)}\;\ee
Using \cite{Chaichian:2000ia} the analogue of the Moyal-Weyl star product of two Weyl symbols $\Phi$ and $\Phi'$ on the cylinder is given by
\be [\Phi\star\Phi'](e^{i \phi},\tau_n)=e^{i\alpha (\partial_ {t'}\partial _{\chi}-\partial_{ t}\partial_{\chi'})}\,\Phi(e^{i( \phi+\chi)},\tau_n+t)\,\Phi'(e^{i (\phi+\chi')},\tau_n+t')\Big|_{t=t'=\chi=\chi'=0}\ee

\bigskip
\newpage
 \appendix
\appendice{\Large{\bf $\quad$  Spinor conventions}}

Here we review our conventions for  spinors in $2+1$ Minkowski space, which are those of \cite{RuizRuiz:1996mm}.  They were utilized in section 5.2.  For the 
$\gamma$-matrices  satisfying $\{\gamma^\mu,\gamma^\nu\}= 2\eta^{\mu\nu}\BI_2$, we utilize the real  $2\times 2$ representation
\be  [\gamma^\mu]^\alpha_{\;\;\beta} =\{-i\sigma_2,\sigma_1,\sigma_3\} \;,\label{Aone}\ee where $\sigma_i$ are the Pauli matrices and   $\alpha,\beta,...=1,2$ denote the spinor indices.  The gamma matrices act on two-dimensional spinors $\psi^\alpha$.  Spinor indices are raised and lowered with $\epsilon_{\alpha\beta}$:  $\psi_\alpha=\psi^\beta\epsilon_{\beta\alpha}$, $\psi^\alpha=\epsilon^{\alpha\beta}\psi_\beta$.  From the representation (\ref{Aone}), it follows  in  that $ [\gamma^\mu]_{\alpha\beta}$ and $ [\gamma^\mu]^{\alpha\beta}$ are symmetric matrices.  Majorana spinors $\psi^\alpha$ satisfy $\bar\psi=\psi^\dagger i\gamma^0= \psi^T C$, where $\dagger$ and $T$ denote the adjoint and transpose, respectively, and $C$ is the charge conjugation matrix, satisfying  $C\gamma^\mu C^{-1}=-\gamma^{\mu\,T}$.   All real spinors are Majorana since we can take $C=i\gamma^0$. 
For any pair of  real  Majorana spinors $\psi$ and $\chi$  one has the identities
\be  \bar \psi\chi =\bar \chi \psi\qquad\quad \bar \psi\gamma^\mu\chi =-\bar \chi \gamma^\mu\psi\label{btwo}\ee
Additional identities for the gamma matrices are 
\beqa
 \gamma^\mu\gamma^\nu&=&-\epsilon^{\mu\nu\rho}\gamma_\rho +\eta^{\mu\nu}\BI_2\label{bthree}\\
\gamma^\mu\gamma^\nu\gamma^\rho&=&-\epsilon^{\mu\nu\rho}\BI_2
-\eta^{\mu\rho}\gamma^\nu +\eta^{\nu\rho}\gamma^\mu+\eta^{\mu\nu}\gamma^\rho\label{bfour}
\eeqa

\bigskip
{\Large {\bf Acknowledgments} }

\noindent
I am very grateful to A. Chaney, L. Lu, A.  Pinzul and S. Sarker for valuable discussions.

\end{document}